# Distribution Fitting 1. Parameters Estimation under Assumption of Agreement between Observation and Model

**Lorentz JÄNTSCHI**

Technical University of Cluj-Napoca, 400641 Cluj, Romania; http://lori.academicdirect.org

**Abstract.** The methods for parameter estimation under assumption of agreement between observation and model are reviewed. The distribution parameters are obtained for one set of experimental data by using different estimation methods under assumption of Gauss-Laplace theoretical distribution. The results are presented and discussed.

## INTRODUCTION

Let be $Y = (Y_1, ..., Y_n)$ and $X = (X_1, ..., X_n)$ series of pair observations and the objective be finding of a function $f(x; a_1, ..., a_m)$ for which $\hat{Y} = f(X)$ is the best possible solution of the approximation $\hat{Y} \sim Y$. Reaching this objective suppose finding of the expression of the $f$ function and of the values of $a_1, ..., a_m$ parameters. Under assumption of agreement between observation and model the expression of the function $f$ is supposed to be known (or at least supposed, when a search from a given set of alternative expressions is conducted). Thus, it is remaining to obtain the values of $a_1, ..., a_m$ parameters. In order to have a unique solution for the values of the $a_1, ..., a_m$ parameters at least $m \leq n$ is required to be assured. A series of alternatives are available for $Y \sim f(X)$ approximation, that ones considered most important being revised and exemplified in this research.

## MATERIALS AND METHODS

1. *Minimizing the error of agreement (minimizing the disagreement)*. Under this assumption a series of alternatives are available (Eq(1), different $p$ and $q$). When the series $Y$ represents (not null) frequencies of the distinct observations $X$ then $f(x)$ should be a positive function too, and the modulus in numerator of (2) and it is no longer required. Assumption of Gauss distribution (Gauss, 1809) of the values of the terms under summation in Eq(1)-(3) are translated into $p = 2$, and assumption of Laplace distribution (Laplace, 1812) are translated into $p = 1$ (Fisher, 1920). The probability density function (PDF) of some representatives of the family containing standard $(\mu=0, \sigma=1)$ Gauss and Laplace distributions are exemplified in Figure 1. Minimizing the error of agreement for different $p$ values give different solutions for the parameters, and as can be seen in Figure 1, are associated with different error shapes. Two particular cases are commonly used to estimate the unknown parameters of a distribution when $p = 2$ (Fisher and Mackenzie, 1923; Fisher, 1924). The most general approach to obtain the distribution parameters is to guess the values or to apply an iterative procedure which reduces in every step the quantity given by the Eq(1) until the reduced quantity is much less than the reminded one.

$$S(p,q) = \sum_{i=1}^{n} |Y_i - f(X_i)|^p / f^q(X_i) = \min., \quad q = 0, 1, \,^p/_2, \, p \tag{1}$$

$$GL(0;0.5) = \sqrt{15/2} \cong 2.739$$
$$GL(0;1.0) = \sqrt{1/2} \cong 0.707$$
$$GL(0;2.0) = 1/\sqrt{2\pi} \cong 0.399$$
$$GL(0;3.0) = ... \cong 0.342$$
$$GL(0;4.0) = \Gamma^2(3/4)\cdot 2^{1/4} \cdot \pi^{-3/2} \cong 0.321$$

$$GL(x;p) = \frac{p}{2} \frac{\Gamma^{1/2}(3/p)}{\Gamma^{3/2}(1/p)} \exp\left(-\frac{|x|^p}{\left(\frac{\Gamma(1/p)}{\Gamma(3/p)}\right)^{p/2}}\right)$$

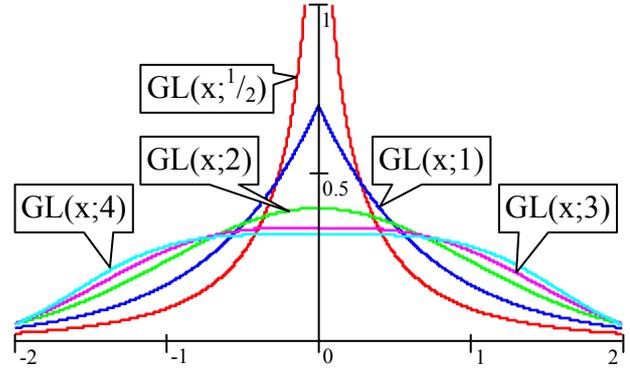

Figure 1. Family of distributions having Gauss and Laplace as representatives

2. <u>Using moments</u>. Under assumption that $Y \sim f(X)$ should be even more accurate (second assumption being the randomness of the error with a zero mean) the approximations is used $\Sigma X_i^k Y_i \sim \Sigma X_i^k \cdot f(X_i)$ for $k \geq 0$. The moments method give maximum weight to the first moments, thus a solution $a_1, ..., a_m$ of $Y \sim f(X)$ may come from Eq(2) (the most convenient way for the general case is the iteration of the $a_1, ..., a_m$ parameters staring from some guess values):

$$\sum_{i=1}^{n} X_i^k Y_i \sim \sum_{i=1}^{n} X_i^k f(X_i), \; k = 0, 1, \ldots \qquad (2)$$

3. <u>Using central moments</u>. The $Y \sim f(X)$ assumption strongest the approximations $\Sigma X_i Y_i \sim \Sigma X_i \cdot f(X_i)$ and $\Sigma(Y_i - \overline{Y})^k \sim \Sigma(X_i - \overline{X})^k \cdot f(X_i)$ for $k \geq 2$. The moments method give maximum weight to first central moments, thus a solution $a_1, ..., a_m$ of $Y \sim f(X)$ may come from Eq(3).

$$\sum_{i=1}^{n} X_i Y_i \sim \sum_{i=1}^{n} X_i f(X_i) \text{ and } \sum_{i=1}^{n}(Y_i - \overline{Y})^k \sim \sum_{i=1}^{n}(X_i - \overline{X})^k f(X_i), \; k = 2, 3 \ldots \qquad (3)$$

4. <u>Using population statistics</u>. A slight modification of the previous method may benefit from the availability of population mean, standard deviation, skewness and kurtosis expressions depending on distribution parameters for a large number of well known distributions. For example the μ (mean) and σ (standard deviation) estimated parameters for a normal distributed population come from Eq(4).

$$\mu = \sum_{i=1}^{n} X_i Y_i \Big/ \sum_{i=1}^{n} Y_i \text{ and } \sigma^2 = \left(\sum_{i=1}^{n}(X_i Y_i - \mu)^2\right) \Big/ \left(\sum_{i=1}^{n} Y_i - 1\right) \qquad (4)$$

Thus, by using this simple reasoning, if the theoretical distribution has m parameters ($a_1, ..., a_m$), then to obtain a solution for it is necessary to know the expression of the first $m$ central moments and then to solve the equations like (4) relating population statistics with their estimators from sample.

5. <u>Maximum likelihood estimation</u>. The principle of the maximum likelihood is that a reasonable estimate for a parameter is the one which maximizes the probability (P in Eq(5)) of obtaining the experimental data (Fisher, 1912), and the most probable set of $a_1, \ldots, a_m$ parameters will make P a maximum (and then MLE is maximum).

$$\text{MLE} = \log_2(P) = \sum_{i=1}^{n} \log_2(f(X_i)) \qquad (5)$$

Application. One set of experimental data were taken from literature in order to illustrate the procedures described above (Jäntschi and others, 2009). The measurements were for octanol water partition coefficient ($K_{ow}$) for 205 out of 206 polychlorinated biphenils expressed in logarithmic scale ($\log_{10}(K_{ow})$, $[K_{ow}]=1$). Maximum likelihood estimation was applied for estimation of $\log(K_{ow})$ of investigated PCB's. The Grubbs test was applied in order to identify the outliers (). One experimental data was considered to be an outlier and was not included in estimation of distribution. Table 1 contains the experimental data in ascending order.

Table 1. Two data sets of measurements under assumption of normal distribution

| $\log(K_{ow})$ for 206 polychlorinated biphenils (sorted data) |
|---|
| 4.151; 4.401; 4.421; 4.601; 4.941; 5.021; 5.023; 5.15; 5.18; 5.295; 5.301; 5.311; 5.311; 5.335; 5.343; 5.404; 5.421; 5.447; 5.452; 5.452; 5.481; 5.504; 5.517; 5.537; 5.537; 5.551; 5.561; 5.572; 5.577; 5.577; 5.627; 5.637; 5.637; 5.667; 5.667; 5.671; 5.677; 5.677; 5.691; 5.717; 5.743; 5.751; 5.757; 5.761; 5.767; 5.767; 5.787; 5.811; 5.817; 5.827; 5.867; 5.897; 5.897; 5.904; 5.943; 5.957; 5.957; 5.987; 6.041; 6.047; 6.047; 6.047; 6.057; 6.077; 6.091; 6.111; 6.117; 6.117; 6.137; 6.137; 6.137; 6.137; 6.137; 6.142; 6.167; 6.177; 6.177; 6.177; 6.204; 6.207; 6.221; 6.227; 6.227; 6.231; 6.237; 6.257; 6.267; 6.267; 6.267; 6.291; 6.304; 6.327; 6.357; 6.357; 6.367; 6.367; 6.371; 6.427; 6.457; 6.467; 6.487; 6.497; 6.511; 6.517; 6.517; 6.523; 6.532; 6.547; 6.583; 6.587; 6.587; 6.587; 6.607; 6.611; 6.647; 6.647; 6.647; 6.647; 6.647; 6.657; 6.657; 6.671; 6.671; 6.677; 6.677; 6.677; 6.697; 6.704; 6.717; 6.717; 6.737; 6.737; 6.737; 6.747; 6.767; 6.767; 6.767; 6.797; 6.827; 6.857; 6.867; 6.897; 6.897; 6.937; 6.937; 6.957; 6.961; 6.997; 7.027; 7.027; 7.027; 7.057; 7.071; 7.087; 7.087; 7.117; 7.117; 7.117; 7.121; 7.123; 7.147; 7.151; 7.177; 7.177; 7.187; 7.187; 7.207; 7.207; 7.207; 7.211; 7.247; 7.247; 7.277; 7.277; 7.277; 7.281; 7.304; 7.307; 7.307; 7.321; 7.337; 7.367; 7.391; 7.427; 7.441; 7.467; 7.516; 7.527; 7.527; 7.557; 7.567; 7.592; 7.627; 7.627; 7.657; 7.657; 7.717; 7.747; 7.751; 7.933; 8.007; 8.164; 8.423; 8.683; 9.143; ~~9.603~~ |

The experimental data were subject of the analysis of the agreement between observation and model by using the following methods: minimizing the error of agreement, and maximum likelihood estimation.

## RESULTS AND DISCUSSION

The Gauss-Laplace distribution general form (from Figure 1) and the kurtosis depending on *p* are given in Figure 2 (Skewness being null).

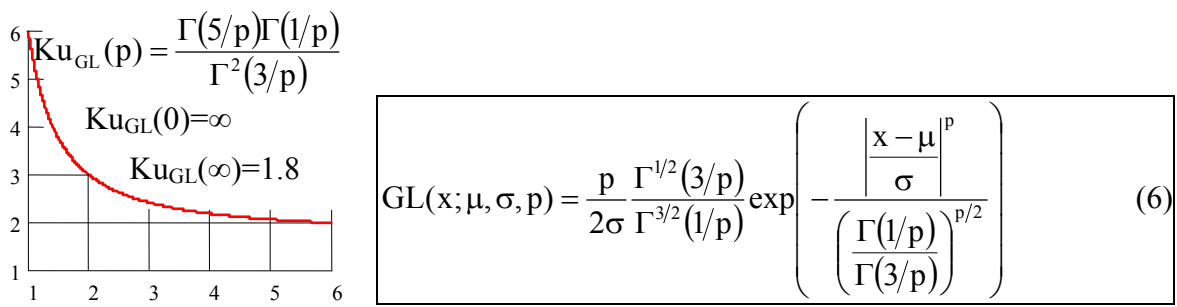

Figure 2. Gauss-Laplace distribution and its kurtosis

$$Ku_{GL}(p) = \frac{\Gamma(5/p)\Gamma(1/p)}{\Gamma^2(3/p)}$$

$Ku_{GL}(0)=\infty$

$Ku_{GL}(\infty)=1.8$

$$GL(x;\mu,\sigma,p) = \frac{p}{2\sigma}\frac{\Gamma^{1/2}(3/p)}{\Gamma^{3/2}(1/p)}\exp\left(-\frac{\left|\frac{x-\mu}{\sigma}\right|^p}{\left(\frac{\Gamma(1/p)}{\Gamma(3/p)}\right)^{p/2}}\right) \qquad (6)$$

The Table 2 contains the estimations of the mean (*μ*) and standard deviation (*σ*) parameters for the data sets given in Table 1. The Eq(1) was used when the minimization of the error of agreement was applied. The Eq(5) was used when the maximum likelihood estimation was applied. The values presented in Table 2 were obtained for different *p* values (0.5-6). The 3D representation of mean, standard deviation and coefficient of variation when minimization of the error of agreement is investigated are graphically represented in Figure 3. The results obtained by maximum likelihood estimation where graphically represented in Figure 4.

Table 2. Mean and standard deviation estimators obtained with Eq(1) and Eq(5)

| Eq. | p | μ | σ | Residues | Eq. | p | μ | σ | Residues |
|---|---|---|---|---|---|---|---|---|---|
| Minimization the error of agreement | | | | | Minimization the error of agreement | | | | |
| Eq(1) q=0 | 0.5 | 6.638 | 1.454 | 128.595 | Eq(1) q=$^p/_2$ | 2.0 | 6.380 | 0.832 | 384.000 |
| Eq(1) q=0 | 1.0 | 6.646 | 0.822 | 128.197 | Eq(1) q=$^p/_2$ | 2.5 | 6.364 | 0.840 | 661.000 |
| Eq(1) q=0 | 1.5 | 6.580 | 0.814 | 143.966 | Eq(1) q=$^p/_2$ | 3.0 | 6.356 | 0.862 | 1250.000 |
| Eq(1) q=0 | 2.0 | 6.544 | 0.748 | 176.138 | Eq(1) q=$^p/_2$ | 3.5 | 6.384 | 0.891 | 2570.000 |
| Eq(1) q=0 | 2.5 | 6.512 | 0.722 | 231.718 | Eq(1) q=$^p/_2$ | 4.0 | 6.400 | 0.916 | 5643.000 |
| Eq(1) q=0 | 3.0 | 6.512 | 0.698 | 323.083 | Eq(1) q=$^p/_2$ | 6.0 | 6.440 | 1.000 | $2 \cdot 10^5$ |
| Eq(1) q=0 | 3.5 | 6.512 | 0.684 | 472.428 | Eq(1) q=p | 6.0 | 6.400 | 1.046 | $4 \cdot 10^7$ |
| Eq(1) q=0 | 4.0 | 6.512 | 0.676 | 720.191 | Eq(1) q=p | 4.0 | 6.350 | 0.960 | $2 \cdot 10^5$ |
| Eq(1) q=0 | 6.0 | 6.472 | 0.644 | 5204.000 | Eq(1) q=p | 3.5 | 6.314 | 0.946 | $5 \cdot 10^4$ |
| Eq(1) q=1 | 6.0 | 6.446 | 0.906 | 17510.000 | Eq(1) q=p | 3.0 | 6.298 | 0.930 | 16230.000 |
| Eq(1) q=1 | 4.0 | 6.430 | 0.856 | 1646.000 | Eq(1) q=p | 2.5 | 6.312 | 0.936 | 5431.000 |
| Eq(1) q=1 | 3.5 | 6.422 | 0.840 | 1025.000 | Eq(1) q=p | 2.0 | 6.328 | 0.952 | 1924.000 |
| Eq(1) q=1 | 3.0 | 6.398 | 0.816 | 678.400 | Eq(1) q=p | 1.5 | 6.352 | 0.984 | 747.800 |
| Eq(1) q=1 | 2.5 | 6.388 | 0.814 | 485.600 | Eq(1) q=p | 1.0 | 6.386 | 1.102 | 344.739 |
| Eq(1) q=1 | 2.0 | 6.380 | 0.830 | 383.890 | Eq(1) q=p | 0.5 | 6.588 | 1.924 | 205.386 |
| Eq(1) q=1 | 1.5 | 6.388 | 0.896 | 340.000 | | | | | |
| Eq(1) q=1 | 1.0 | 6.388 | 1.104 | 344.700 | Maximum likelihood estimation | | | | |
| Eq(1) q=1 | 0.5 | 6.366 | 2.112 | 427.100 | Eq(5) | 4 | 6.476 | 0.886 | -373.810 |
| Eq(1) q=$^p/_2$ | 0.5 | 6.650 | 1.904 | 156.800 | Eq(5) | 3 | 6.468 | 0.829 | -360.790 |
| Eq(1) q=$^p/_2$ | 1.0 | 6.486 | 0.970 | 185.800 | Eq(5) | 2 | 6.464 | 0.802 | -354.208 |
| Eq(1) q=$^p/_2$ | 1.5 | 6.412 | 0.856 | 249.000 | Eq(5) | 1 | 6.510 | 0.914 | -371.620 |

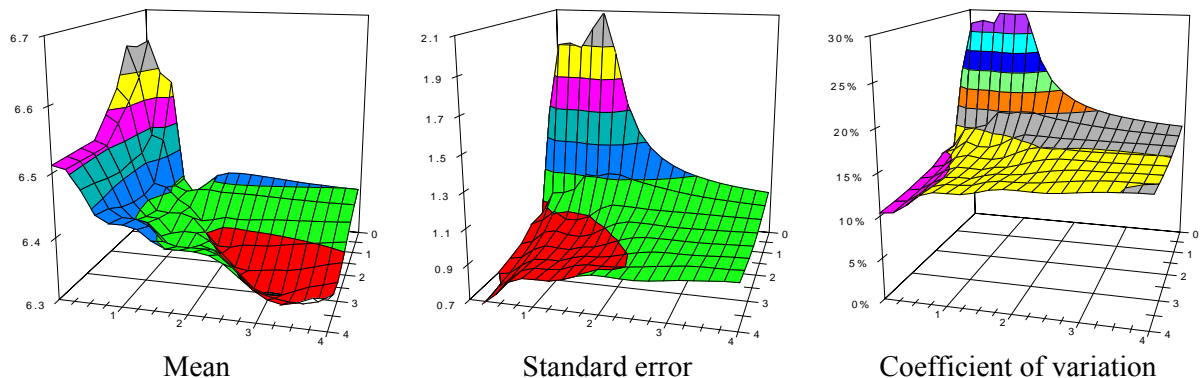

Mean          Standard error          Coefficient of variation

Figure 3. Minimizing the error of agreement: variation of statistics for different *p* and *q* values
(p - from Gauss-Laplace distribution and Eq(1), q - from Eq(1))

The analysis of the results obtained by using the minimization agreement errors approach (see Table 2 and Figure 3) leads to the following remarks:
- The evolution of μ for 0.5 ≤ p ≤ 3 is similar for q=p/2 and q=p as expected.
- The evolution of σ for p is similar for q=1, p=p/2 and q=p.
- The minimum amplitude for a given p value is obtained for q=1 when the μ is analyzed and for q=0 when the σ is analyzed.
- The same trend in variation of μ is observed for p=0.5, p=1, p=1.5, p=2 (up-down-up-down). The highest variation was between q=1 and q=p/2 for p=0.5. The mean values decrease as q increase for a given value of p≥2.5.
- The highest values of σ were observed for p=0.5 with a pick for q=1. The same behaviour but with smaller differences were observed for p=1, p=1.5, p=2. A similar behaviour of σ variation was observed for p≥2.5; the smaller variation is observed for p=2.5 (the smaller difference of standard deviation as the *q* value increased).

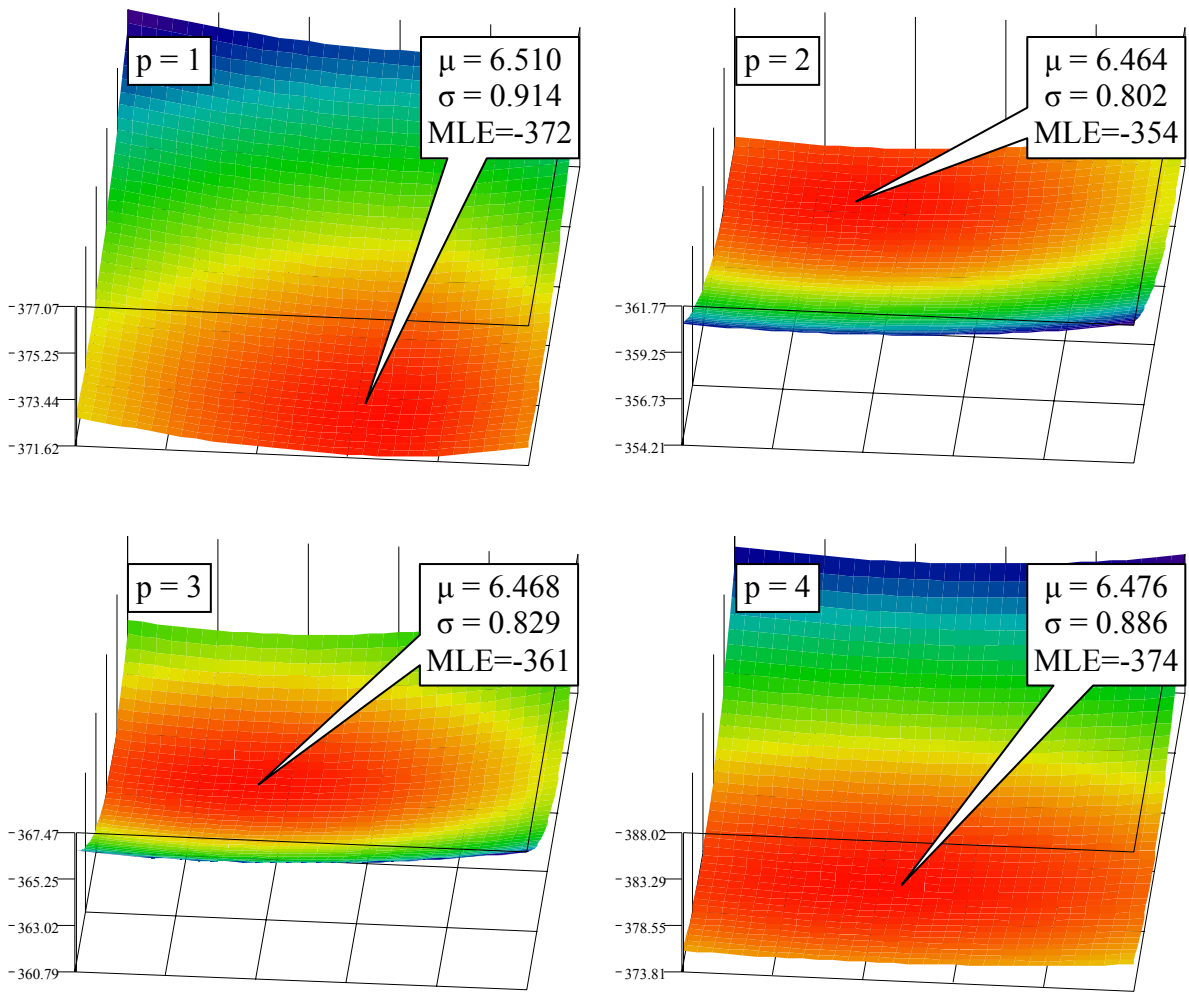

Figure 4. Likelihood estimation and its maximum (MLE) for different *p* values
(μ - mean, σ - standard deviation, p - from Gauss-Laplace distribution)

The analysis of the results obtained by using the maximum likelihood estimation approach (see Table 2 and Figure 4) leads to the following remarks:
- The *μ* and *σ* varied slightly with increases of *p* value (an amplitude of 0.046 was obtained for *μ* and of 0.112 for *σ*);

- The maximum value of $\mu$ and $\sigma$ were obtained for p=1;
- The minimum value of $\mu$ and $\sigma$ were obtained for p=2;
- A decrease of $\mu$ and $\sigma$ is observed for p=2. Starting with p=2 their values increase slightly with increases of *p* values.

It may be concluded that as *p* increases the minimizing of the error of the agreement give more weight to the outliers (see Figure 5).

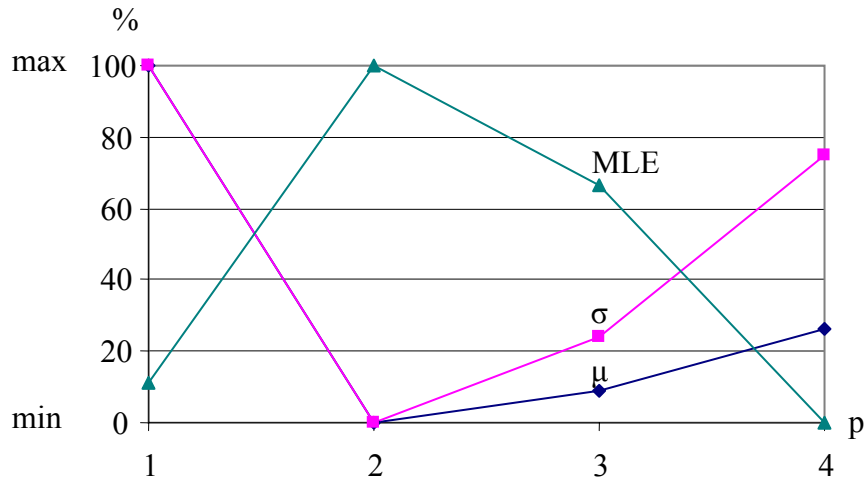

Figure 5. Mean ($\mu$), standard deviation ($\sigma$) and maximum likelihood (MLE) on a relative scale (between min and max) for p = 1, 2, 3, 4 (Gauss-Laplace distribution)

The obtained MLE estimation is presented in Figure 6. The associated equation and its statistical characteristics are presented in Eq(9).

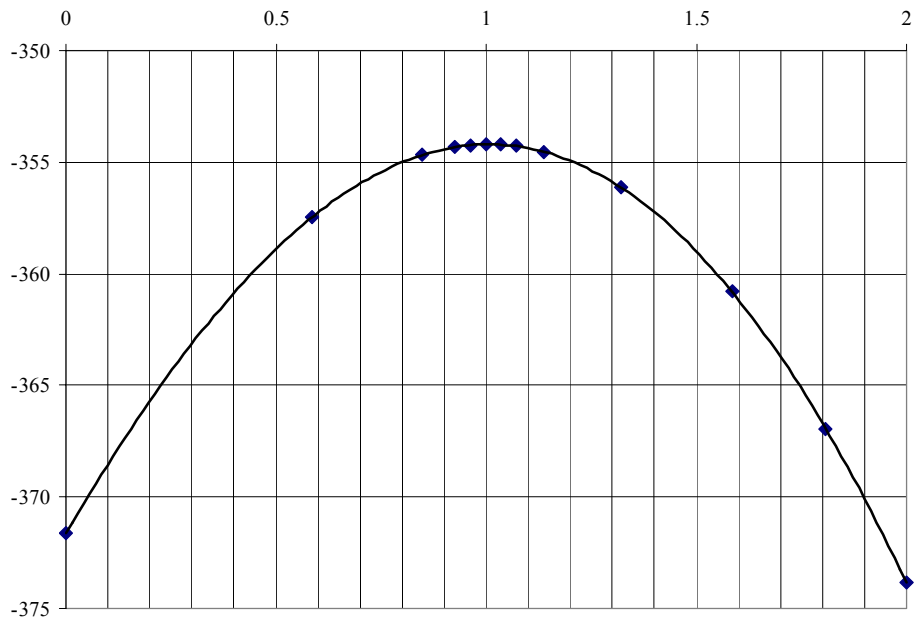

Figure 6. Maximum likelihood estimation for distribution of log($K_{ow}$) with 205 of 206 PCBs

$$MLE = .585 \cdot (\log_2(p))^4 - 3.67 \cdot (\log_2(p))^3 - 11.6 \cdot (\log_2(p))^2 + 32.1 \cdot \log_2(p) - 372$$
$$r^2 > 0.99999; \quad p_{max} = 2.008; \quad MLE_{max} = -354.207$$
(9)

Eq(9) shows that with a great accuracy the maximum likelihood (MLE=log(P), Eq(5)) of the Gauss-Laplace distribution (GL(x,μ,σ,p), Eq(6)) can be approximated by a fourth order polynomial formula on log(p).

For normal (or Gauss) distributed data (as the investigated dataset, the common case of the biochemical data) the maximum of this polynomial is expected to be near p = 2 and for error function (or Laplace) distributed data (the common case of astrophysical data) the maximum of this polynomial is expected to be near p = 1. From this point of view, the observed data shows a very good agreement with the normal distribution, maximum likelihood of the Gauss-Laplace distribution being estimated at p = 2.008.

CONCLUSIONS

Five methods of parameter estimation under assumption of agreement between observation and model were reviewed. The abilities of minimizing the error of agreement and maximum likelihood estimation were applied on a set of PCBs experimental data. The results showed that as *p* increases the minimizing of the error of the agreement give more weight to the outliers. As maximum likelihood estimation is concerned, a powerful model in terms of estimation was obtained when p = 2.008.

*Acknowledgments.* The research was partly supported by UEFISCSU Romania through ID1051/202/2007 grant, PN2 (2007-2010) national research framework.